\documentclass[
preprint,
amsmath,
amssymb,
aps,
pra,
floatfix,
]{revtex4-1}

\usepackage{graphicx}
\usepackage{dcolumn}
\usepackage{bm}
\usepackage{siunitx}
\usepackage{epstopdf}
\usepackage{xcolor}
\usepackage{hyperref}

\begin{document}


\title{Vacuum-Ultraviolet Absorption and Emission Spectroscopy of Gaseous, Liquid, and Supercritical Xenon}

\author{Christian Wahl}
\email{wahl@iap.uni-bonn.de}
\author{Marvin Hoffmann}
\author{Thilo vom Hoevel}
\author{Frank Vewinger}
\author{Martin Weitz}
 \email{weitz@uni-bonn.de}
\affiliation{%
 Institut für Angewandte Physik, Universität Bonn\\
 Wegelerstraße 8, 53115 Bonn, Germany
}%

\date{\today}

\begin{abstract}
Bose-Einstein condensation, an effect long known for material particles as cold atomic gases, has in recent years also been observed for photons in microscopic optical cavitites.
Here, we report absorption and emission spectroscopic measurements on the lowest electronic transition ($5\text{p}^6 \rightarrow 5\text{p}^5 6\text{s}$) of xenon, motivated by the search for a thermalization medium for photon Bose-Einstein condensation in the vacuum-ultraviolet spectral regime.
We have recorded pressure-broadened xenon spectra in the \SIrange{135}{190}{\nano\metre} wavelength regime at conditions near the critical point.
The explored pressure and temperature range includes high pressure gaseous xenon below the critical pressure and supercritical xenon at room temperature, as well as liquid xenon close to the boiling point near the critical pressure.
\end{abstract}

                             
\maketitle

\section{Introduction}
Bose-Einstein condensation has been experimentally observed with ultracold atomic gases, exciton polaritons, and more recently with photons, e.g. in dye solution-filled optical microcavities \cite{einstein1914quantentheorie, cornell2002nobel, ketterle2002nobel, anderson1995observation,kasprzak2006bose, deng2010exciton, balili2007bose, klaersbose2010, marelic2015experimental, greveling2018density}.
In the latter system, a short mirror spacing in the wavelength regime provides a low-frequency cutoff, and visible spectral range photons confined in the resonator thermalize to the dye, which is at room temperature, by repeated absorption re-emission processes. 
The dye molecules fulfill the thermodynamic Kennard-Stepanov relation between absorption and emission spectral profiles. 
When photons in the cavity thermalize faster than they are lost through e.g. mirror transmission, Bose-Einstein condensation to a macroscopically occupied ground state is observed \cite{klaersbose2010, kirton2013nonequilibrium, schmitt2015thermalization}. 
Other than in the well-known laser, no inverted active medium is required and spontaneous emission is retrapped. 
Thus, the usual argument of ultraviolet coherent sources showing disfavourable pump power scaling and thus making it hard to operate at high optical frequencies here does not apply. 
In previous work, we have proposed photon condensation as a possible means to realize coherent optical sources in the UV spectral regime, and discussed the use of noble gases at high pressure as possible thermalization media with closed electronic transitions from the ground to the lowest electronically excited state lying in the vacuum ultraviolet spectral regime (VUV) \cite{wahl2016absorption}. 
Frequent collisions of the gas atoms can lead to a thermalization of quasimolecular states, which is a prerequisite for the Boltzmann-like Kennard-Stepanov scaling between absorption and re-emission in this gaseous system \cite{moroshkin2014kennard, Kennard, Stepanov, PhysRevA.54.4837}. 
A further important prerequisite for thermalization of UV photons by repeated absorption re-emission processes of course is a sufficient spectral overlap between absorption and emission spectral profiles, such that the emission can be reabsorbed by the gas atoms during the cavity lifetime.
In a binary collisional model (and when neglecting the effect of pressure shifts), the molecular xenon emission line has a Stokes shift of \SI{25}{\nano\metre} with respect to the \SI{147}{\nano\metre} absorption line. 

The motivation of the present work is to obtain data of absorption and emission spectral profiles in dense xenon samples, which is a regime where spectral data on this system is limited.
In earlier works studying VUV absorption of xenon in the gaseous regime, broadband measurements up to \SI{30}{\bar} pressure at room temperature have been reported \cite{wahl2016absorption}. 
For monochromatic light at \SI{173}{\nano\metre} wavelength at a pressure approaching the critical pressure of \SI{58.4}{\bar} complete absorption in a \SI{13}{\centi\metre} long cell has been reported, indicating strong absorption far off-resonant from the $5\text{p}^6 \rightarrow 5\text{p}^5 6\text{s}$ transition centered at \SI{147}{\nano\metre} wavelength \cite{koehler1974vacuum}. 
We are not aware of other published measurements for supercritical xenon in absorption or emission spectroscopy, despite its wide use in chromatography, and as a non-linear medium for temporally compressing ultra-short laser pulses \cite{guiochon2011fundamental,didenko2018compressor}.

Xenon in the liquid state is to date used in large scale scintillation detectors in the search for dark matter and neutrino research \cite{aprile2017xenon1t, baldini2005absorption}.
For this use mainly two properties are of interest - namely the absorption length for the scintillation light emitted by xenon and the accurate wavelength of this emitted light.
For the absorption length in the literature two contradicting results are available. 
Modern measurements find an attenuation length for the scintillation emission for liquid xenon at close to \SI{-110}{\celsius} temperature of up to \SI{100}{\centi\metre}, with the value being mainly determined by impurities in the xenon \cite{baldini2005absorption, solovov2004measurement,grace2017index}.
On the other hand, an early broadband absorption measurement of liquid xenon close to its boiling point, which at pressures of approximately \SI{50}{\bar} is a few degrees centigrade below room temperature, reported a complete absorption of light of wavelengths up to \SI{226.5}{\nano\metre} through a cell of several millimeter length \cite{mclennan1930absorption}.
Since the two measurements were recorded at different temperatures, they do not directly contradict each other although such a large difference in absorption seems hard to justify theoretically.
Work studying the emission spectrum of xenon in the liquid phase has reported the emission to be centred around \SI{178\pm 0.6}{\nano\metre} \cite{jortner1965localized} or \SI{174}{\nano\metre} \cite{doke1999present}, respectively, while the most recent report by Fujii et al. reports a center wavelength of \SI[parse-numbers=false]{174.8 \pm 0.1 (stat) \pm 0.1 (sys.)}{\nano\metre} \cite{fujii2015high}.

Here we report measurements of the absorption and emission of the $5\text{p}^6 \rightarrow 5\text{p}^5 6\text{s}$ electronic transition of xenon. 
While earlier work of our group has studied absorption spectra in the gaseous regime up to \SI{30}{\bar} pressure at room temperature, here we extend the investigated pressure range up to \SI{130}{\bar}, thus well reaching into the supercritical regime. 
We have moreover recorded data for samples cooled to the liquid phase. 
Besides the described absorption spectral measurements, also emission spectral data for the described phases are reported.

In the following, chapter II describes the used experimental setup and chapter III contains absorption measurements of xenon samples. 
In the subsequent chapter IV, emission measurements are reported, and finally chapter V gives conclusions.

\section{Experimental setup}
\label{sec:Experimental setup}
\begin{figure*}[htbp!]
 \resizebox{12.7cm}{!}{%
 \includegraphics{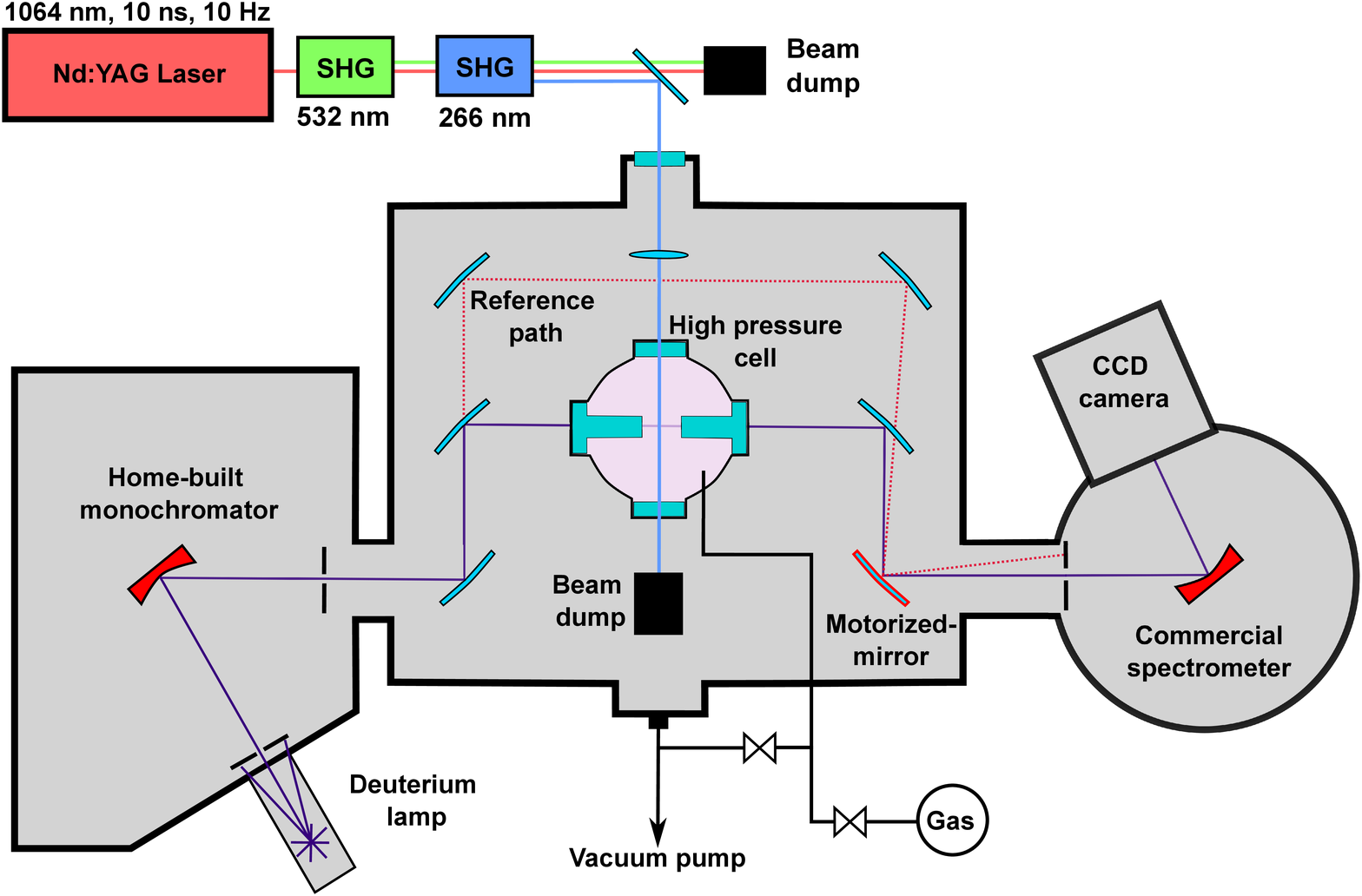}
 }
\caption{Schematic representation of the experimental setup used for absorption and emission spectroscopy. 
For absorption measurements light emitted by a deuterium lamp is spectrally filtered by a constant deviation monochromator, passes through a high pressure cell containing the sample, and is then spectrally analysed. 
Part of the light is split off before reaching the sample and guided around the high pressure cell to monitor the output power of the deuterium lamp.
Using a motorized mirror light from one of the two paths is directed into the detection spectrometer. 
To perform emission spectroscopic measurements, light of the fourth harmonic of a pulsed Nd:YAG laser is sent in an orthogonal direction into the high pressure cell. 
Light being emitted near an angle of \SI{90}{\degree} can then be analysed using the same optical path as used for the absorption measurement. 
}
\label{fig:Setup}
\end{figure*}

A schematic representation of the experimental setup used to analyse vacuum-ultraviolet absorption and emission properties of xenon is shown in Fig.\,\ref{fig:Setup}.
A high pressure cell containing xenon is placed inside a vacuum chamber which is evacuated to \SI{5e-5}{\milli\bar} to supress absorption of light of a wavelength of less than \SI{193}{\nano\metre} by atmospheric oxygen and nitrogen molecules.

The absorption spectroscopy setup is similar as previously reported \cite{wahl2016absorption}.
In brief, for absorption spectroscopy a \SI{150}{\watt} water-cooled deuterium lamp (Hamamatsu L1835) is used as a light source, permitting spectroscopic measurements down to a wavelength of \SI{115}{\nano\metre}.
Given that the deuterium lamp emission spectrum has large variations of the emission spectral density (near two orders of magnitudes over the spectral regime of interest), a double monochromator setup is used. 
A home-built first monochromator utilizes a curved grating with \num{1200} groves/mm, with its blaze angle optimized for \SI{200}{\nano\metre} wavelength radiation in a constant deviation monochromator scheme. 
With a \SI{1}{\milli\metre} wide exit slit the width of the produced spectral bandpass is \SI{5.5}{\nano\metre}.

The transmitted radiation passing the first monochromator is subsequently split up, with one part passing through the high-pressure cell containing the gas to be analyzed and the second one being sent around the cell to act as a reference path, allowing for normalization of the spectroscopic signal. 
The radiation passing the cell enters a second monochromator, for which a commercial device (McPherson 234/302 \SI{200}{\milli \meter}) is used.
This monochromator is equipped with a 2400 groves/mm grating.
For a slit width of \SI{10}{\micro \meter} a wavelength resolution of \SI{0.05}{\nano\meter} is reached.
The resulting wavelength-resolved signal is detected using a cooled open nose VUV sensitive CCD camera.
A flip mirror placed before the detection monochromator allows for path selection between the reference path and the path through the spectroscopic pressure cell.
The used cell is constructed to sustain gas pressures up to \SI{200}{\bar}. 
We use a metal cell with MgF$_2$ windows. 
On the optical axis used for absorption measurements the cell is equipped with step windows, such that the optical path length in the cell can be tuned by using windows of a suitable step height.
In the presented measurements an effective cell length of \SI{180}{\micro \meter} on the optical axis is used in the absorption measurements.

For spectroscopic measurements of the emission, a different excitation light source is used.
Here, a Nd:YAG laser (Quanta-Ray GCR-12S) is utilized, which produces \SI{10}{\nano\second} long pulses near \SI{1064}{\nano\metre} wavelength with a repetition rate of \SI{10}{\hertz} and a pulse energy of \SI{250}{\milli \joule}.
With the fourth harmonic of this laser near \SI{266}{\nano\metre} wavelength and an optical pulse energy of \SI{10}{\milli \joule}, we can excite the pressure-broadened $5\text{p}^6 \rightarrow 5\text{p}^5 6\text{p}$ xenon two-photon transition, which has a transition wavelength of \SI{128}{\nano\metre}, off-resonantly. 
It is interesting to note that at higher xenon densities a second two-photon excitation channel near \SI{133}{\nano\metre} transition wavelength arises, which in earlier work has been attributed to the formation of excitons \cite{pournasr1996even}. 
This second channel can be excited almost resonantly with the used irradiation wavelength.
The pump light near \SI{266}{\nano\metre} is focussed into the high pressure cell with a lens of \SI{20}{\centi\metre} focal length along a direction perpedicular to the path utilized for absorption spectroscopy.
In this way, the same analysis optics and spectrometer as described above can be used to record the time integrated spectrally resolved fluorescence and phosphorescence signal of the xenon sample.

For the part of the measurements using gas pressures above a pressure of used xenon gas bottle pressure (near \SI{56}{\bar}), a manual spindle press (Sitec 750.1030) is used to compress the xenon gas.
To allow for spectroscopic measurements of xenon in the liquid phase, a cooling of the cell is required. 
For this, the xenon cell can be equipped with a cooling jacket, through which liquid nitrogen is fed at a variable flow rate. 
The reachable cell temperature of $\approx \SI{-20}{\celsius}$ are clearly below the critical temperature of xenon (\SI{16.58}{\celsius}), thus spectroscopic measurements in the liquid phase can well be carried out with the present setup.

\section{Xenon VUV absorption measurements}

\begin{figure}[htbp!]
\resizebox{8.6cm}{!}{%
 \includegraphics{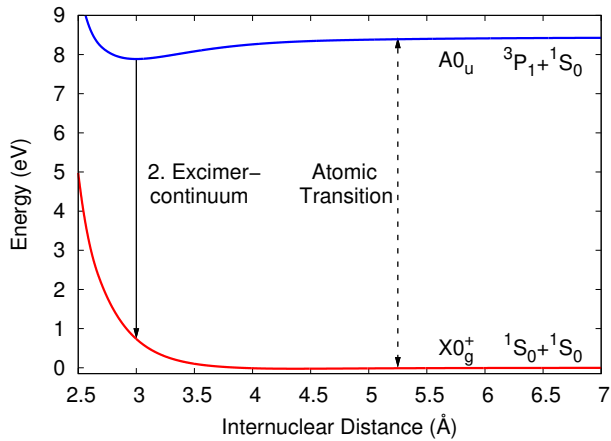}
 }
\caption{Calculated energy curves for the ground sate X$0^+_g$ and the lowest lying electronically excited state A$0_u$ in the xenon dimer system versus the internuclear distance \cite{selg2003visualization}.}
\label{fig:Lines}
\end{figure}

\begin{figure}[htbp!]
\resizebox{8.6cm}{!}{%
 \includegraphics{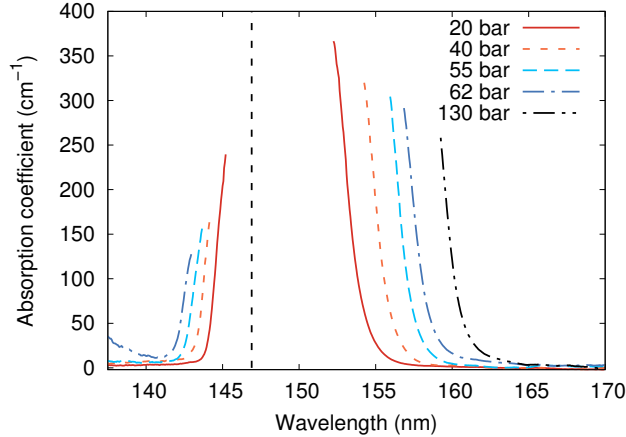}
 }
\caption{Room temperature ($T \approx 1.04 T_c$ where $T_c = \SI{16.6}{\celsius}$ denotes the critical temperature of xenon) absorption spectra of xenon in the wavelength range of \SIrange{137.5}{170}{\nano\metre} for pressures between \SIlist{20;130}{\bar}.
Saturated regions are omitted for clarity.
The vertical dashed line gives the position of the line center of the $5\text{p}^6 \rightarrow 5\text{p}^5 6\text{s}$ transition in the absence of pressure broadening and shift.
}
\label{fig:Measurement_Gas}
\end{figure}

\begin{figure}[htbp!]
\resizebox{8.6cm}{!}{%
 \includegraphics{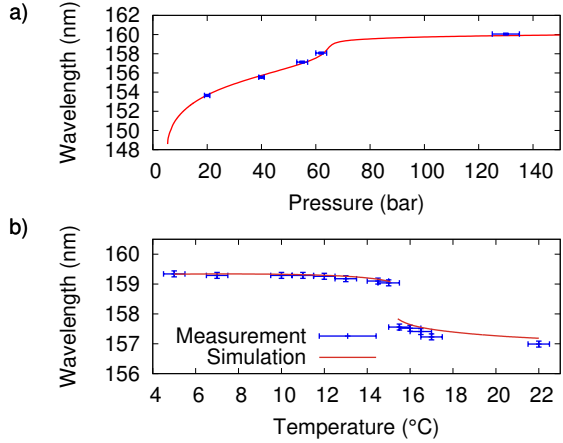}
 }
\caption{(a) Experimentally determined wavelength $\lambda_{\SI{100}{\per\centi\metre}}$ at which the absorption coefficient on the red wing xenon $5\text{p}^6 \rightarrow 5\text{p}^5 6\text{s}$ resonance reaches a value of \SI{100}{\per\centi\metre} (data points) along with theory (solid red line) versus the gas pressure at room temperature.
(b) Variation of $\lambda_{\SI{100}{\per\centi\metre}}$ on the cell temperature, for a constant pressure of \SI{57}{\bar} ($P \approx 0.98 P_c$, 
where $P_c = \SI{58.4}{\bar}$ denotes the critical pressure of xenon).
The observed jump at the wavelength $\lambda_{\SI{100}{\per\centi\metre}}$ occurs at a temperature consistent with the liquid to gaseous phase transition temperature corresponding to \SI{15.38}{\celsius} at the used pressure value.
}
\label{fig:MeasSim}
\end{figure}

To begin with, absorption spectra of the xenon $5\text{p}^6 \rightarrow 5\text{p}^5 6\text{s}$ electronic transition have been recorded, i.e. the transition from the ground state to the lowest-lying electronically excited state, in the wavelength range around \SI{147}{\nano\metre}.
From the ground and singly excited potential energy curves for two colliding xenon atoms, see Fig. \ref{fig:Lines}, it can be seen that for not to small internuclear distances (above \SI{4}{\angstrom}), when the ground state energy curve still is mostly flat, the (singly) excited state energy curve already bends to lower energies for decreasing internuclear distance. 
The described shape of the potential curves is the origin of the significantly stronger broadening of the red wing of the pressure-broadened absorption line in comparison to the blue wing in a regime beyond the impact limit at the used high gas pressures \cite{borovich1973pressure}.
Fig. \ref{fig:Measurement_Gas} shows pressure broadened xenon VUV absorption spectra observed for different values of the gas pressure. 
Besides the $5\text{p}^6 \rightarrow 5\text{p}^5 6\text{s}$ transition (centered at \SI{147}{\nano\metre} wavelength in the thin vapor regime) on the short wavelength side also the red wing of the next higher energetic xenon line, centered at \SI{129.5}{\nano\metre} wavelength, is observed starting with the data recorded at \SI{62}{\bar} pressure. 
For the data recorded at the highest used pressure (\SI{130}{\bar}) the two absorption line signals already overlap.
Near the line center, the $5\text{p}^6 \rightarrow 5\text{p}^5 6\text{s}$ experimental absorption data is saturated, with corresponding data being here omitted for clarity. 
We observe a strong pressure broadening and shift to the red side of the resonance, the latter understood from the already above mentioned enhancement of the red wing originating from the form of the quasimolecular potential curves. 
A more detailed inspection of the red wing of the absorption profile, for which also at higher pressures no overlap with other lines is present, shows that the observed line shape far of resonance here well matches an exponential decay. 
This observation remains valid throughout the gaseous and supercritical phases of the xenon gas (above and below near \SI{58.4}{\bar} respectively) at the used room temperature conditions.
In order to compare the individual measurements, in the following the wavelength at which the red wing of the absorption profile reaches an absorption coefficient of \SI{100}{\per\centi\metre} is evaluated and denoted as $\lambda_{\SI{100}{\per\centi\metre}}$.
The corresponding quantity is relevant as to judge whether the enhanced red wing results in a reabsorption of the emission spectrum of the transition despite the large (\SI{25}{\nano\metre} in the absence of pressure broadening) Stokes shift.
For a gas-filled microcavity with e.g. \SI{1.5}{\micro\metre} mirror spacing, fluorescent radiation of wavelength $\lambda_{\SI{100}{\per\centi\metre}}$ would be reabsorbed following 30 cavity round trips, which is in the order of magnitude of the realizable cavity finesse values with dielectic mirrors in this wavelength range.
Table \ref{tab:Meas-Table} summarizes the obtained results for room temperature conditions and the data points in Fig. \ref{fig:MeasSim}a display the corresponding variation with cell pressure, showing an increase of the wavelength shift from the unperturbed line center at which an absorption coefficient of \SI{100}{\per\centi\metre} is reached. 

\begin{table*}[htbp]
\begin{center}
\begin{tabular}{| c | c | c | c |}
\hline
Pressure (\si{\bar}) & Density (\si{\kilogram\per\cubic\metre}) & $\lambda_{\SI{100}{\per\centi\metre}}$ \text{[red wing]} (\si{\nano\metre}) & $\lambda_{\SI{100}{\per\centi\metre}}$ \text{[blue wing]} (\si{\nano\metre}) \\
\hline
\num{20.0\pm 0.1} & \num{122 \pm 1} &\num{153.65 \pm 0.16} &\num{144.45 \pm 0.16}\\
\num{40.0\pm 0.1} & \num{297 \pm 1} & \num{155.55 \pm 0.16}&\num{143.75 \pm 0.16}\\
\num{55.0\pm 0.5} & \num{540 \pm 13} & \num{157.15 \pm 0.16}&\num{143.15 \pm 0.16}\\
\num{62.0\pm 0.5} & \num{977 \pm 107} & \num{158.05 \pm 0.16}&\num{142.60 \pm 0.16}\\
\num{130.0\pm 5} & \num{1988 \pm 16} & \num{160.05 \pm 0.16}& N/A \\
\hline
\end{tabular}
\end{center}
\caption{Measured values for the wavelength at which the absorption coefficient of the spectrum of the lowest energetic xenon absorption line at room temperature reaches a value of \SI{100}{\per\centi\metre} on the red and the blue wing of the resonance.
}
\label{tab:Meas-Table}
\end{table*}

To allow for a comparison of this data with theory, we have carried out numerical simulations.
These numerical simulations are based on a calculation of the Franck-Condon factors for the system \cite{selg2003visualization}. 
For the simulation, it is assumed that the absorption on the wing of the profile is mainly due to the effect of two-body collisions, and the quasistatic approximation is used \cite{hedges1972extreme,moe1976absorption,allard1982effect,alioua2006classical}.
The influence of multiparticle collisions, which can be relevant at the used high densities, should be mostly visible close to the resonance wavelength of the observed transition. 
Thus higher order collisions are not considered in the numerical simulation and the whole profile can be derived from the potential energy curves in Fig. \ref{fig:Lines}.

The red solid line in Fig. \ref{fig:MeasSim}a shows the result of our simulations for the expected value of $\lambda_{\SI{100}{\per\centi\metre}}$ versus pressure. 
Our experimental results are in good agreement with the calculations, which supports the assumption that the far red wing of the absorption profile is governed by the effect of two-body collisions.
At the highest experimentally investigated pressure of \SI{130}{\bar} xenon exhibits a density of \SI{1988}{\kilo\gram\per\cubic\metre}.
A further increase in pressure up to \SI{4194}{\bar}, the pressure at which xenon at room temperature would become solid \cite{linstrom2001nist}, would lead to a further increase of the density by a factor \num{1.7}, and, following our simulations, lead to an additional red shift of $\lambda_{\SI{100}{\per\centi\metre}}$ by only about \SI{1.1}{\nano\metre}.
We however point out that at such densities, which approach those of a solid, the influence of multiparticle collisions should be carefully evaluated, which may well result into more significant corrections to the line shape than in our presently experimentally accessible parameter regime.

In further measurements, we have recorded corresponding data for different cell temperatures, which at the used constant gas pressure of \SI{57}{\bar} allows both the gaseous and, upon cooling to below a temperature of \SI{15.38}{\celsius} (corresponding to the expected value of this phase transition at this pressure value), the liquid phase to be analyzed. 
Fig. 4b shows the variation of the wavelength $\lambda_{\SI{100}{\per\centi\metre}}$ versus the cell temperature (data points) along with the dependence expected from our simulation (red solid line). 
Upon cooling to below the phase transition temperature, we observe a shift of $\lambda_{\SI{100}{\per\centi\metre}}$ to the red by about \SI{1.1}{\nano\metre}, with the experimental results being in good agreement with our theory predictions. 
The corresponding increased width of the absorption line in the red wing is well understood by the density increase (here corresponding to a factor of about \num{1.7}) upon entry into the liquid phase.
The results are in stark contrast to previous measurements by McLennan and Turnbull who under comparable conditions found a complete absorption of light up to a wavelength of \SI{226.5}{\nano\metre} through a cell of several \si{\milli\metre} length \cite{mclennan1930absorption}.
It is, however, consistent with previous findings for liquid xenon at a temperature of \SI{-108}{\celsius}, where an absorption length for light at \SI{178}{\nano\metre} of more than \SI{100}{\centi\metre} was observed \cite{baldini2005absorption}.

\section{Xenon emission measurements}

We have next carried out spectroscopic measurements of the vacuum-ultraviolet emission of gaseous xenon. 
At pressures above \SI{100}{\milli\bar}, upon excitation of the lowest energetic electronically excited of the xenon atom, the $5\text{p}^5 6\text{s}$ state, it is well known that Stokes-shifted radiation centered at near \SI{172}{\nano\metre} wavelength is emitted, the so-called second excimer continuum \cite{borovich1973pressure,castex1981experimental,sieck1968continuum,brodmann1978xenon,borovich1975contribution}. 
This is well understood from the potential curves of the xenon dimer system (Fig. \ref{fig:Lines}), with the singly excited state being bound thus forming an excimer system.
Accordingly, when exciting the $5\text{p}^6 \rightarrow 5\text{p}^5 6\text{p}$ two-photon transition off-resonantly, the emission is found spectrally peaked around the second excimer continuum at \SI{172}{\nano\metre} wavelength. 
We attribute de-excitation from the $5\text{p}^5 6\text{p}$ state to the $5\text{p}^5 6\text{s}$ state in the used high-pressure regime to be dominated by collisional processes given the occurring multiple crossings of the quasimolecular electronically excited states curves \cite{bowering1986collisional}.
In our experiment, we monitor the emission from the subsequently expected decay to the electronic ground state, which is in the vacuum-ultraviolet.

\begin{figure}[htbp!]
\resizebox{8.6cm}{!}{%
    \includegraphics{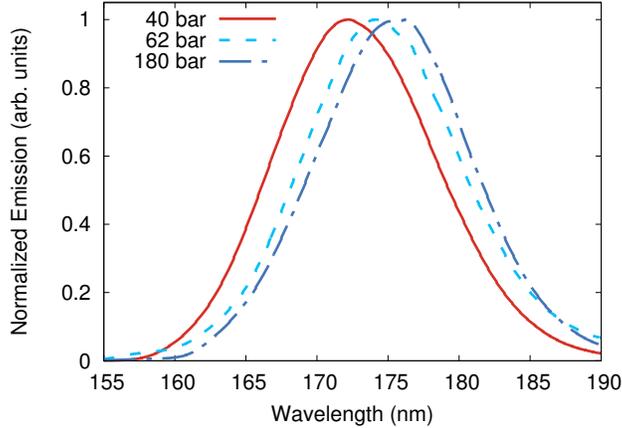}
    }
\caption{Measured spectra of the xenon second excimer continuum for three different values of the xenon pressure at room temperature ($T = 1.04 T_c$).}
\label{fig:Fluo}
\end{figure}

In Fig. \ref{fig:Fluo} spectrally resolved emission measurements recorded at room temperature for three different xenon pressures are presented. 
All three spectra are normalized to their respective maximum value for better comparability since the excitation efficiency of the used two-photon transition depends strongly on the xenon density.
For gaseous xenon at \SI{40}{\bar} pressure the peak of the emission spectrum is found at \SI{172.2\pm 0.1}{\nano\metre}, which agrees within experimental uncertainties with previous measurements of the emission of gaseous xenon \cite{mccann1988two,sieck1968continuum,borovich1975contribution}.
The two spectra shown by blue dashed lines, recorded at \SI{62}{\bar} and \SI{180}{\bar} pressure respectively, corresponding to pressure values above the critical pressure, are clearly shifted to longer wavelengths.
For the highest pressure data (with \SI{180}{\bar} pressure), the maximum of the emission is centered at \SI{175.3 \pm 0.1}{\nano\metre} wavelength, corresponding to a red shift of \SI{3.1}{\nano\metre} with respect to the \SI{40}{\bar} data. 
The full width of the emission spectrum for the three shown spectra remains nearly constant at \SI{13.3\pm 0.2}{\nano\metre}. 
Given that the observed width stays nearly constant despite the large density difference in spectra, we conjecture that the observed red shift of the supercritical and liquid phase emission data with respect to the gas emission is not due to reabsorption effects.

\begin{figure*}[htbp!]
\resizebox{12.9cm}{!}{%
    \includegraphics{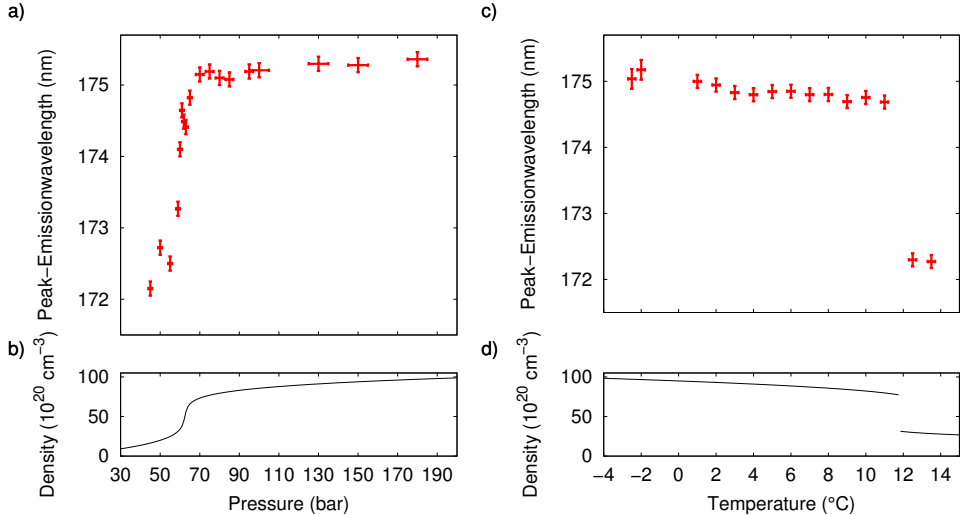}
    }
\caption{a) Observed variation of the spectral maximum of the xenon second excimer continuum emission versus the xenon pressure recorded at room temperature ($T \approx 1.04 T_c$), see b) for the dependence of the xenon density on pressure.
c) Gives corresponding experimental data for the variation of the peak emission with temperature recorded at the constant pressure of \SI{53}{\bar} ($P \approx 0.91 P_c$), see d) for the dependence of xenon density on temperature. 
Here the phase transition temperature is $T \approx \SI{11.8}{\celsius}$.}
\label{fig:6}
\end{figure*}

Fig. \ref{fig:6}a shows the variation of the peak emission wavelength versus pressure. 
The observed red shift coincides with the increase of the gas density upon pressure variation, see Fig. \ref{fig:6}b. 
In the supercritical regime, a continuous variation of density with gas pressure occurs. 
Given that the used room temperature conditions are only a few degrees above the critical temperature of xenon, the density increase at values near the critical pressure visible in Fig. \ref{fig:6}b is relatively rapid, which leads to the fast increase of the pressure shift in this regime observed in Fig. \ref{fig:6}a.
Fig. \ref{fig:6}c shows data for the variation of the peak emission wavelength on temperature, recorded at the constant value of the pressure above the liquid of \SI{53}{\bar}; see Fig. 6d for the corresponding variation of density. 
The density here exhibits a discontinuity upon cooling from the gaseous to the liquid phase, given that the used pressure is below the critical pressure (phase transition temperature: \SI{11.8}{\celsius} at a pressure of \SI{53}{\bar}). 
The observed sudden variation of the peak of the emission visible in Fig. \ref{fig:6}c near this temperature is in good agreement with the expected discontinuous density change upon crossing the phase transition from liquid to gaseous. 
In order to compare our results to earlier work, we average our available data both above and below the phase transition temperature, yielding results for the peak emission wavelength of the second eximer continuum of \SI{174.9 \pm 0.4}{\nano\metre} and \SI{172.0 \pm 1.0}{\nano\metre} wavelength in the liquid and gaseous phase of xenon respectively.
Other works report the emission of liquid xenon to be centred around \SI{178\pm 0.6}{\nano\metre} \cite{jortner1965localized} or \SI{174}{\nano\metre} \cite{doke1999present}, while the here performed measurement most closely agrees with the recent work of Fujii et al. with a reported center wavelength of \SI[parse-numbers=false]{174.8 \pm 0.1 (stat) \pm 0.1 (syst.)}{\nano\metre} \cite{fujii2015high}.

\section{Conclusions}
To conclude, absorption and emission spectroscopic measurements of the xenon $5\text{p}^6 \rightarrow 5\text{p}^5 6\text{s}$ electronic transition at densities close to the critical point are reported. 
In the absorption measurements, in agreement with e.g. the lower pressure regime data of \cite{borovich1973pressure}, we observe an asymmetric lineshape with an enhanced red wing. 
Data for the far red wing has been compared to numerical simulations based on two-body collisions also described in our work, which gives good agreement with our experimental results even at densities up to \SI{2000}{\kilogram\per\metre\cubed}.
In contrast to earlier measurements by Koehler et al. for supercritical xenon \cite{koehler1974vacuum}, as well as by McLennan and Turnbull for liquid xenon \cite{mclennan1930absorption} no significant broadening of the absorption profile extending above what can be expected from the binary potential energy curves was observed. 

In our measurements, other than in earlier work, xenon VUV emission spectra have also been observed in the supercritical phase.
It was shown that the peak emission wavelength of supercritical xenon at room temperature exhibits a similar shift compared to the dilute gas emission as observed in liquid xenon. 
As applies to the liquid phase case, also for the supercritical xenon case at high pressure, the observed shift of the peak emission wavelength of in both cases near \SI{3}{\nano\metre} cannot be explained by the binary potential energy curves, based on which no change in the peak emission wavelength would be expected.
Future theory work here should analyse for the possible influence of multiparticle collisions.
Within the supercritical phase, the magnitude of the observed shift of the peak emission wavelength may be utilized to monitor whether the sample is more in a "gas-like" or a "liquid-like" state, for small and large spectral shifts of the emission respectively.
This can provide a spectroscopic means to determine the position of the so-called Widom-line in supercritical xenon \cite{mcmillan2010going, simeoni2010widom}.

In the investigated pressure and temperature regimes, the pressure shifts reduce the Stokes shift between absorption and emission not yet by an amount to allow for sufficient reabsorption enabling the thermalization of photons by absorption and re-emission processes as in visible spectral range dye microcavity experiments \cite{klaersbose2010, klaers2011bose}. 
For the future, the influence of manybody collisional effects on the lineshapes should be investigated, which can be relevant for higher gas pressures than experimentally investigated in the present work. 
A different possible approach is to investigate the influence of higher gas temperatures, as numerical simulations predict a sufficient spectral overlap at temperatures of \SI{1000}{\kelvin} and above.
The increased possible reabsorption of fluorescent radiation is understood from that ground state atoms at such elevated temperatures during diatomic collisions can climb up to smaller internuclear distances in the quasimolecular potential curves (Fig. \ref{fig:Lines}) than at room temperature.
Metal-coated multilayer mirrors \cite{pazidis2013optical} can allow for the realization of suitable microcavities in this temperature regime, where the use of dielectic cavity mirrors is not feasible.
Once sufficient spectral overlap is established, it will be important to test for the Boltzmann-like Kennard-Stepanov scaling between absorption and re-emission profiles. 
Other attractive candidates for photon gas thermalization in the vacuum-ultraviolet regime are mixtures of small molecules that are gaseous at room temperatures as the CO molecule with a light noble gas acting as a buffer gas, i.e. which itself is transparent in the spectral regime of the molecular lines. 

We acknowledge support of the Deutsche Forschungsgemeinschaft within SFB/TR 185 (277625399).

\bibliographystyle{spphys}
\bibliography{wahl_refs}
\end{document}